\documentclass[journal]{IEEEtran}
\usepackage{graphicx}
\begin{document}

\title{Multifractal age? \\ Multifractal analysis of  cardiac interbeat
intervals in  assessing of  healthy aging}

\author{Danuta Makowiec, Stanis\l aw Kryszewski, \\ Joanna Wdowczyk-Szulc,
Marta \.Zarczy\'nska-Buchowiecka, Marcin Gru\-cha\-{\l a}, Andrzej Rynkiewicz,
\thanks{Danuta Makowiec and Stanis\l aw Kryszewski  are  with
Gda\'nsk University, Poland,  e-mail: fizdm@univ.gda.pl.}
\thanks{Joanna Wdowczyk-Szulc, Marta \.Zarczy\'nska-Buchowiecka,
Marcin Gru\-cha\-{\l a} and Andrzej Rynkiewicz are with Gda\'nsk
Medical University, Poland. }
}

\markboth{Proceedings of the 7{\tiny th} ESGCO 2012, April 22-25, 2012,
Kazimierz Dolny, Poland}%
{Makowiec \MakeLowercase{\textit{et al.}}: Multifractal age? \\
Multifractal analysis of  cardiac interbeat intervals in assessing of
healthy aging}

\maketitle


\begin{abstract}
24-hour Holter recordings of 124 healthy people at different age are
studied.  The nocturnal signals of young people reveal the presence
of  the multiplicative structure. This structure is significantly
weaker in diurnal signals and becomes less evident for elderly people.
Multifractal analysis allows us to propose  qualitative and quantitative
methods to estimate the advancement of the aging process
for healthy humans.
\end{abstract}



\section{Introduction}
The structure-function-based multifractal analysis relies on scaling
properties of some statistical measure of the data set (see  \cite{Gao}).
Namely, if $\{ X_i, ~i=1,2,\dots, N \} $ is a time series for which
the multifractality is investigated, and  $R(i,n)$ is a function
that measures  a certain property of a signal in the $i$th box
containing $n$ consecutive points of data (boxes do not overlap),
then the multifractal analysis  considers scaling properties of all real
value moments of  $R(i,n)$. Then, the question is whether the following
power-law dependence  on box size $n$ exists or not for various real $q$'s:
\begin{equation}
      Z(n,q)= < |R(i,n)|^q >_{\{X_i\}} \sim n^{\tau(q)},
\label{eq01} \end{equation}
where $<.>_{\{X_i\}}$ denotes averaging over the corresponding
data sets. When it is found that such a power-law scaling is present
for some $q$, then we say that the studied process has fractal
structure. Furthermore, if the scaling exponent function $\tau(q)$
is not linear in $q$ then it is said that the process is
multifractal. The multifractal spectrum of such a process: $h\rightarrow D(h)$ is obtained from the scaling exponent function $\tau(q)$ by the Legendre
transformation $(q,\tau(q))\rightarrow (h,D(h))$.

Different values of $q$ correspond to certain statistical
properties of a signal. For example, when $q=2$, the method provides
estimates for the Hurst exponent, $H= (\tau(2) +1)/2$.
If $q\rightarrow \infty$ then we get information about large changes in the signal, while  $q\rightarrow -\infty$ characterizes the smooth parts of a series.
These properties support and motivate our efforts in developing methods based on the multifractal approach.

It often appears that the scaling curves, namely  $\log Z(n,q)$ vs $\log n$
plots are, at least at the first sight, linear. Hence, according
to the above given arguments, the signal can be classified as
fractal. However, more careful investigation allows one to distinguish
different scaling regimes. One type of scaling is present for the
short-time scales and another one for the long ones.
Such variations of scaling can be related to changes in the intrinsic
properties of the considered time series \cite{Stoev, Kantelhardt2009}.

The cardiac interbeat time intervals, so called RR-signals, are studied in four  generally accepted  \cite{TASK}  frequency regimes:
\begin{itemize}
\item high-frequency (HF) $(0.15, 0.4)$~Hz,
\item low-frequency (LF) $(0.04,0.15)$~Hz,
\item very-low-frequency (VLF)  $(0.0033,0.04)$~Hz,
\item ultra-low-frequency (ULF)  $<0.0033 $~Hz.
\end{itemize}
Such a division is associated with different, physiologically
distinguishable  aspects of cardiac rhythm control. Therefore,
by considering time scales  in  the scaling procedure  
corresponding to the listed  frequency bands, we hope to get insights
into  these aspects.  In practical application, assuming that the mean cardiac
interbeat is 0.8 s, we have assigned  size of "boxes" for investigation of the
scaling properties in signals. In particular, we assume   $n = 32,\dots, 420$  RR-intervals as corresponding  to the VLF band

Multifractal analysis based on the study of structure functions must be
applied with care. If the data exhibits noise properties then the
partial summation must be performed before  the proper analysis
\cite{Gao}. On the other hand, if the data has some features of the
random walk, there is no need for partial summation. Therefore,
before multifractal analysis one needs to decide whether the data
should be treated as a noise process or as a random walk. In the
following we will show what additional information can be gained
if the initial data is considered both as a noise process and as
a random walk.

The method summarized above was applied to 24-hour Holter recordings
of cardiac time interbeat signals obtained from healthy persons.
The obtained results led to the estimators which allowed
us to describe and (at least to some extent) quantify the advance
of the aging process. The specific details are given in
\cite{Makowiec2009, Makowiec2011a, Makowiec2011b}.
Extensive discussion, numerous examples are presented in just cited
articles. Here we focus only on the essentials.

\section{Methods}

24-hour Holter monitoring of ECG was performed for 124 healthy subjects:
{\em young adults}: 36 persons (age 18-26),
{\em middle-aged adults}: 40 persons (age 45-53),
{\em elderly}: 48 persons (age 65-94).
Thus, we have analyzed RR-signals consisting of time intervals between
consecutive normal cardiac contractions which originated in the sinus node.

Multifractal analysis, similarly as in the time-domain statistics,
is found to be robust to data removal (see \cite{Clifford}).
However, multifractal statistics requires long consecutive series 
to have access to large pool of scales. Additionally, studied series 
should have a stable character.
Therefore, we carefully divided the obtained signals into two parts
corresponding to nocturnal and diurnal (daily) activities.
The high levels of  meanRR, SDNN and pNN50 were assumed to specify 
sleep hours. For each RR-signal the
period of the six consecutive hours with high levels of
these quantities was extracted and labeled as {\em sleep}.
Then, the part corresponding to the typical afternoon activity
was labeled as {\em wake}.

Multifractal analysis relying upon the study of structure functions
was performed with the well-known approach called Wavelet Transform Modulus
Maxima (WTMM) \cite{Muzy}. The WTMM-based analysis was performed
for each person, both for {\em wake} and {\em sleep} parts
of his/her signals. This was done not only for the proper
signal, but also for the corresponding integrated one. The scaling properties were estimated  for scales corresponding to the VLF band.
Then, after the necessary Legendre transformation, the multifractal
spectra were obtained for  {\em wake} and {\em sleep} parts of every
individual RR-signal.

\begin{figure}[h]
\centering
\includegraphics[width=7cm, height=5cm]{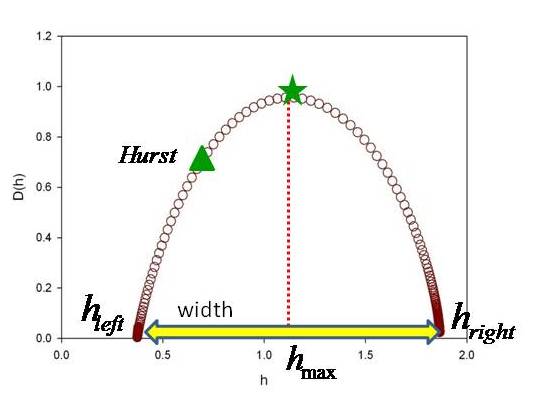}
\caption{Assessment of a typical multifractal spectrum:
$h_{max}$ -- scaling of the most dense subset of a signal;
$Hurst$ -- correlation;
$h_{left}$ -- extreme, and   $h_{right}$ -- smooth events scalings;
$width$ = $h_{right}-h_{left}$ -- variety of subsets with different scalings}
\label{schemat}
\end{figure}

\begin{figure}
\centering
\includegraphics[width=9cm]{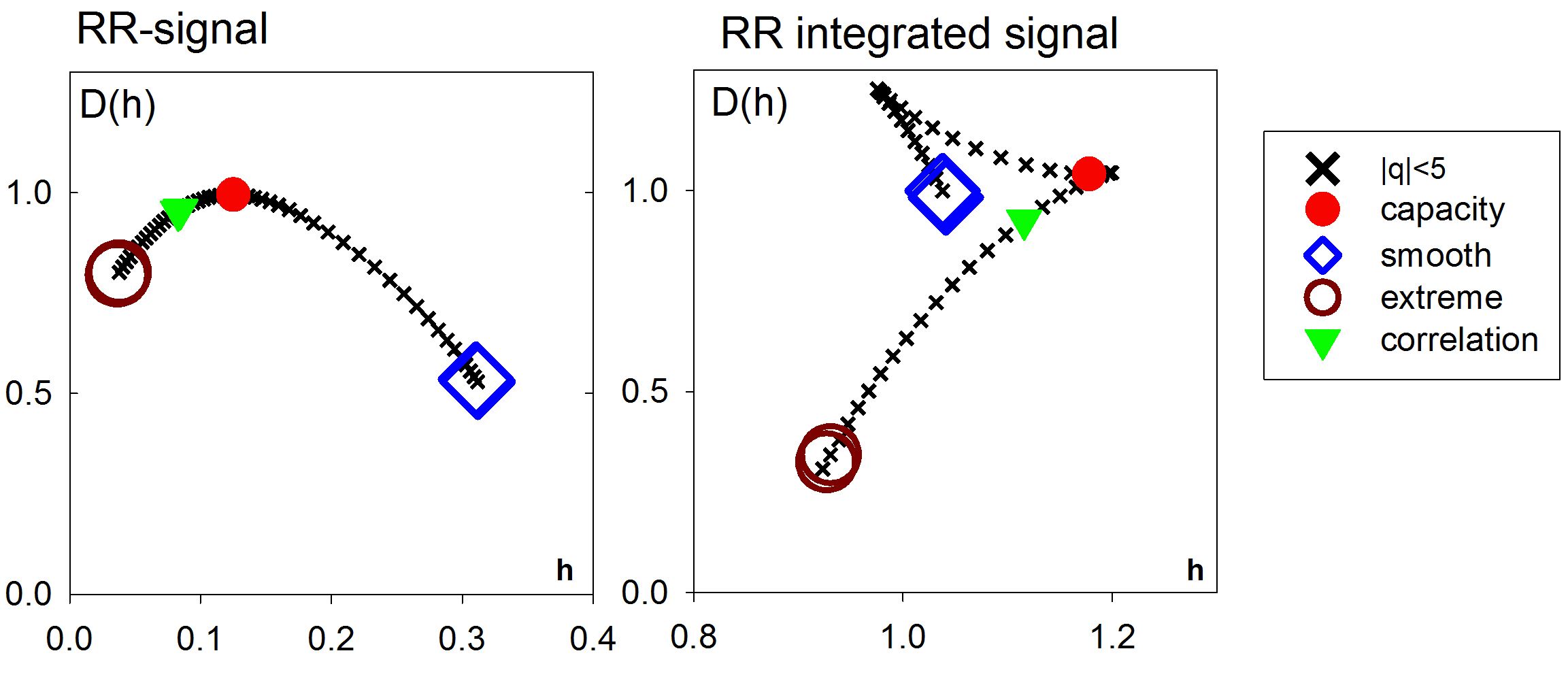}
\caption{Examples of multifractal spectra obtained from RR series.}
\label{examples}
\end{figure}

In Fig.~\ref{schemat} we present the typical multifractal spectrum
and collect the main parameters characterizing such a spectrum.
However, real spectra obtained from cardiac time series differ
considerably from typical ones. Examples of such spectra are presented
in Fig.~(\ref{examples}). The evident irregularities (especially
in the integrated signals) require special care and the corresponding
analysis is quite time-consuming. The details concerning such spectra
(examples and their discussion) can be found in
\cite{Makowiec2011b}.

\section{Results}
\subsection{$\Delta_{max}$ in cardiac  signals}

The distance $\Delta_{max}$ between $h_{max}$ in the spectrum obtained from  the integrated signal and $h_{max}$ in a spectrum obtained in a signal  plays a very important role. It allows to distinguish additive from multiplicative processes.
We have found that such a distinction can be made for diurnal and
nocturnal parts of the signals. The main observation is that the
nocturnal signals  possesses multiplicative character in the VLF region.  Such organization is quite weak in the diurnal signals. Moreover, it was found that
the multiplicative features are less and less pronounced when
the age of the investigated individuals grows. Hence, weakening
of multiplicativity can be considered as an indicator of
aging.

\subsection{Multifractal age}

In our analysis we have tested several parameters to characterize
typical multifractal spectra. In particular, for each person, we have examined:
$h_{max}$,  $H$ (Hurst), $h_{left}$,  $h_{right}$, $\Delta$ --
the width (see Fig.\ref{schemat} for the corresponding illustration) in his\slash her{\it wake} and {\it sleep} parts of the signals and their integrated counterparts.

Tedious statistical analysis for {\em sleep} and {\em wake} parts
of the signals allowed as to formulate the following five
criteria.
\begin{equation}
    h_{max}^{sleep} < h_{max}^{wake},
\label{cr01} \end{equation}
\begin{equation}
    h_{max}^{wake,int} \mbox{is large enough} (> 1.15),
\label{cr02} \end{equation}
\begin{equation}
    \Delta_{max}^{sleep} \approx 1,
\label{cr03} \end{equation}
\begin{equation}
    H^{sleep} < H^{wake},
\label{cr04} \end{equation}
\begin{equation}
    H^{int,sleep} \mbox{is large enough}.
\label{cr05} \end{equation}
The notation was explained above and the additional superscripts
seem to be self-evident. These criteria were applied to all
124 signals obtained from all age groups.

The results are summarized in the table below. The first row tells
us how many positive answers were given to the question whether
each (out of 5) criterion is fulfilled. The main body of the table
presents the percentage of the number of positive answers obtained
for three investigated age groups.
$$ \begin{array}{c|rrrrrr} 
   &\multicolumn{6}{|c}{\mbox{Number of positive answers}} \\
  \mbox{group name: }&0 &1 &2 &3 &4 &5 \\  \hline
  \mbox{young }&53 &25 &11 &11 &0 &0  \\
  \mbox{middle }& 3 & 33 &30 &20 &7 &7  \\
  \mbox{elderly }&0  &10 &10 &21 &29 & 29 \\
\end{array}  $$
The number of positive answers clearly increases with age
and the table has a characteristic skew form.
The skewness of the table contents suggests that we can
associate the number of positive answers to criteria (\ref{cr01})
-- (\ref{cr05}) as indicators of the {\bf multifractal age}.

\section{Conclusions}

We have applied the multifractal analysis to 124 24-hour
RR signals divide into {\em wake} and {\em sleep} parts.
The main conclusions can be summarized as follows
\begin{itemize}
\item The skewness of the presented table stems from
      the careful multifractal analysis of Holter recordings
      of RR signal obtained from healthy individuals.
      The number of positive answers whether the criteria
      are met, clearly increases with age. Thereby, we
      suggest that the term {\bf multifractal age} may
      be useful in the studies of healthy aging.
\item The quantity $\Delta_{max}$ leads to distinction between additivity
      and multiplicativity of the investigated RR signals.
      Hence, we suggest that $\Delta_{max}$ should be accepted
      as an additional parameter characterizing multifractality
      of the time series.

\end{itemize}

\end{document}